\documentclass[conference,10pt]{IEEEtran}	
\usepackage{amsmath,graphicx}
\usepackage{amssymb}
\usepackage{acronym}

\usepackage[belowskip=-10pt,aboveskip=1.8pt]{caption}
\usepackage{standalone}
\usepackage{subcaption}
\usepackage{tikz}
\usepackage{url,cite}
\usepackage{verbatim}
\usepackage[bookmarks,colorlinks]{hyperref}
\usepackage{soul, xcolor}
\usepackage{caption}
\usepackage{subcaption}
\usepackage{colortbl,multirow}
\usepackage[inline]{enumitem}
\usepackage[ruled,linesnumbered,vlined]{algorithm2e}
\SetKwInput{KwData}{\textbf{Init}} 

\newcommand{\myVec}[1]{{\boldsymbol{#1}}}
\newcommand{\myMat}[1]{{\boldsymbol{#1}}}
\newcommand{\mySet}[1]{\mathcal{#1}}
\newcommand{\Ntx}{{N_{\rm tx}}}
\newcommand{\Nrx}{{N_{\rm rx}}}
\newcommand{\Nbins}{{B}}

\newcommand{\Nparams}{{N_{\rm p}}}
\newcommand{\Params}{{\myVec{\varphi}}}
\newcommand{\NoisyParams}{{\myVec{\phi}}}
\newcommand{\DNNParams}{{\myVec{\theta}}}
\newcommand{\SoW}{{\myVec{\psi}}}

\DeclareMathOperator*{\argmax}{arg\,max}

\newtheorem{lemma}{Lemma}

\usepackage[all=normal,paragraphs=tight,floats=normal,mathspacing=normal,wordspacing=tight,charwidths=tight,mathdisplays=normal,leading=normal]{savetrees}
\acrodef{ai}[AI]{artificial intelligence}
\acrodef{dnn}[DNN]{deep neural network}
\acrodef{ris}[RIS]{reconfigurable intelligent surface}
\acrodef{adc}[ADC]{analog to digital convertor}
\acrodef{awgn}[AWGN]{additive white Gaussian noise}
\acrodef{mimo}[MIMO]{multiple-input multiple-output}
\acrodef{mse}[MSE]{mean squared-error}
\acrodef{mmse}[MMSE]{minimal mean squared-error}
\acrodef{map}[MAP]{maximum a-posteriori probability}
\acrodef{los}[LoS]{line-of-sight}
\acrodef{pga}[PGA]{projected gradient ascent}
\acrodef{sgd}[SGD]{stochastic gradient descent}
\acrodef{physfad}[PhysFad]{physics-based end-to-end model of \ac{ris}-parameterized wireless channels with adjustable fading}
\acrodef{snr}[SNR]{signal to noise ratio}
\acrodef{sgm}[SGM]{score-based generative models}
\acrodef{sde}[SDE]{stochastic differential equation}
\acrodef{ald}[ALD]{annealed Langevin dynamics}
\acrodef{zogd}[ZOGD]{zero order gradient descent}
\usetikzlibrary{trees,arrows}

\setstcolor{blue}

\IEEEoverridecommandlockouts

\title{AI-Aided Annealed Langevin Dynamics for Rapid Optimization of Programmable Channels}
\author{
\IEEEauthorblockN{Tomer Shaked, Philipp del Hougne, George C. Alexandropoulos, and Nir Shlezinger\\}
\thanks{ 
    This work was supported by  the European Research Council (ERC) under the ERC starting grant nr. 101163973 (FLAIR). 
    T. Shaked and N. Shlezinger are with the School of ECE, Ben-Gurion University of the Negev, Beer-Sheva, Israel (e-mail: tosha@post.bgu.ac.il, nirshl@bgu.ac.il). 
    P. del Hougne is with Univ Rennes, CNRS, %IETR - UMR 6164 F-35000 
    Rennes, France (e-mail: philipp.del-hougne@univ-rennes.fr).
    G. C. Alexandropoulos is with the  Informatics and Telecommunications Dpt., National and Kapodistrian University of Athens,  Greece, (e-mail: alexandg@di.uoa.gr).
	}
	\vspace{-1.0cm}
 }
\begin{document}

\maketitle
\begin{abstract}
    Emerging technologies such as \acp{ris} make it possible to optimize some parameters of wireless channels. 
    Conventional approaches require relating the channel and its programmable parameters via a simple model that supports rapid optimization, e.g., re-tuning the parameters each time the users move. However, in practice such models are often crude approximations of the channel, and a more faithful description can be obtained via complex simulators, or only by measurements. 
   In this work, we introduce a novel approach for rapid optimization of programmable channels based on AI-aided \ac{ald}, which bypasses the need for explicit channel modeling. %By exploiting the connection between \ac{ald} and diffusion-based generative models, 
   By framing the \ac{ald} algorithm using the \ac{map} estimate,   
   we design a deep unfolded \ac{ald} algorithm that leverages a \ac{dnn} to estimate score gradients for optimizing channel parameters. We introduce a training method that overcomes the need for channel modeling using zero-order gradients, combined with  active learning to enhance generalization, enabling optimization in complex and dynamically changing environments. We evaluate the proposed method in \ac{ris}-aided scenarios subject to rich-scattering effects. Our results demonstrate that our AI-aided \ac{ald} method enables rapid and reliable channel parameter tuning with limited latency.
\end{abstract}

\acresetall

%------------------------------------------------------------------------
%	Introduction
%------------------------------------------------------------------------ 
\section{Introduction}
\label{sec:intro} 
%%%%%%%%%%%%%% Motivation and background         %%%%%%%%%%%%%%
To meet the constant growth in demands for connectivity and coverage, future wireless systems are expected to incorporate various emerging technologies~\cite{giordani2020toward}. These include deploying \acp{ris}  to improve coverage and overcome harsh non-line-of-sight conditions~\cite{ alexandropoulos2021reconfigurable}; using massive \ac{mimo} transceivers; and incorporating novel antenna technologies, based on, e.g., hybrid beamformers~\cite{molisch2017hybrid}, metasurfaces~\cite{shlezinger2021dynamic}, and leaky waveguides~\cite{gabay2023leaky}, to realize massive \ac{mimo} in a cost and power efficient manner. A common characteristic of some of these technologies is their ability to {\em modify the equivalent channel} to some extent. For instance, \acp{ris}  parameterize propagation environments~\cite{faisal2022machine}, while hybrid beamformers and related antennas bring forth the ability to adapt the analog front-end, thus affecting the  end-to-end channel~\cite{shlezinger2023ai}.

%%%%%%%%%%%%%% Literature Survey                 %%%%%%%%%%%%%%
% Model-based optimization
The programmability of  wireless channels enables optimizing the channels and adapting to dynamic variations based on a desired property, e.g., maximizing throughput~\cite{strinati2021reconfigurable}. Common approaches require relating the tunable parameters to the resulting channel via a simple known model. This simple model is then used for conventional parameter optimization via, e.g., first-order  methods~\cite[Ch. 9]{boyd2004convex}, often with limited iterations to support rapid tuning~\cite{lavi2023learn}. However, such simple models are often crude approximations of the environment. For instance, \ac{ris}-aided channel optimization is often based on the cascaded model~\cite{liu2019matrix}. This model was shown to fail to capture various  environments~\cite{rabault2024tacit}, and its formulation requires challenging estimation of the individual channels~\cite{swindlehurst2021channel}, a task  often necessitating dedicated hardware~\cite{alexandropoulos2023hybrid}. 

% Model-free channel optimizaiton
When operating in complex settings,  the tunable parameters can be related to the resulting channel realization only via a complex simulator (e.g., based on ray-tracing~\cite{vitucci2024efficient} or discrete dipole formulations~\cite{faqiri2022physfad}), or by measuring the channel (e.g., via  pilots)~\cite{wang2021jointly}. The absence of a simple closed-form expression relating the channel to the tunable parameters notably complicate their optimization. Existing   model-free approaches based on  Bayesian optimization~\cite{wang2021jointly} or reinforcement learning~\cite{faisal2022machine}, either struggle in tuning many parameters, or tend to be extremely lengthy. 
An alternative approach models complex wireless channels via digital twins, i.e.,  a software replica of a physical environment~\cite{khan2022digital}. In wireless communications, digital twins accommodate both traditional simulators, as well as \ac{ai} models. The latter has been the focus of growing interest in the context of {\em generative models}~\cite{van2024generative}, i.e., to generate~\cite{alkhateeb2023real} and evaluate~\cite{li2023learnable} channels, often employing the popular family of diffusion models~\cite{lee2024generating}. Generative diffusion models were also proposed for channel estimation in cascaded model-based \ac{ris} settings~\cite{zhang2025decision,
tong2024diffusion}. 
The intrinsic relationship between diffusion models and the {\em \ac{ald}} optimizer~\cite{song2019generative} indicates that this \ac{ai} methodology can be adapted from its conventional form of generating samples from complex distributions (i.e., channels), into tackling complex optimization problems~\cite{kawar2022denoising}, motivating its exploration for rapid tuning of programmable channels. %\textcolor{red}{TODO NIR - include reference xxxx that used diffusion models for channel estimation in parameterized channels}

%%%%%%%%%%%%%% Main Contributions                %%%%%%%%%%%%%%
In this work, we study the rapid optimization of programmable channels without relying on simplified models. 
Inspired by the relationship between \ac{ald} optimization and diffusion \ac{ai} models, we formulate an \ac{ai}-aided \ac{ald} algorithm for generating  parameterizations of wireless channels based on the achievable rate measure. Our \ac{ai}-aided \ac{ald} operates with a fixed and limited latency, learning from complex simulations and/or measurements to tune the channel parameters for a given user location.  

We first formulate \ac{ald}  to sample channel parameters based on the achievable rate score. Then, inspired by how diffusion models are obtained from \ac{ald} for image sampling~\cite{song2019generative}, we design a \ac{dnn}  for estimating score gradients, which is used in each \ac{ald} iteration as a form of deep unfolding~\cite{shlezinger2022model}. We introduce a training method that bypasses the need for simple channel modeling by relying on zero-order methods~\cite{duchi2015optimal}. As learning \ac{dnn}-based model for complex channels that generalizes to unseen realizations is often challenging~\cite{stylianopoulos2022deep}, we incorporate active learning tools to boost exploration in training ~\cite{be2019active}. Our experimental study, which considers tuning \ac{ris}-aided channels in rich-scattering environments,  systematically shows that our \ac{ai}-aided \ac{ald}  rapidly finds useful channel configurations.

The rest of this paper is organized as follows: Section~\ref{sec:system model} introduces the channel optimization problem. The proposed method is detailed in Section~\ref{sec:Optimizing Programmable Channels} and numerically evaluated in Section~\ref{sec:simulations}. Finally, Section~\ref{sec:conclusions} provides concluding remarks.

%------------------------------------------------------------------------
%	System Model
%------------------------------------------------------------------------
%\vspace{-0.1cm}
\section{System Model}
\label{sec:system model}
% \vspace{-0.1cm}
% Here, we present the system model. We model the parameterized channel in Subsection~\ref{ssec:paramterized channel}, and formulate the problem of optimizing its configuration in Subsection~\ref{ssec:problem formulation}.

%%%%%%%%%%%%%% Parameterized Channel             %%%%%%%%%%%%%%
%\vspace{-0.1cm}
\subsection{Parameterized Channels}
\label{ssec:paramterized channel}
%\vspace{-0.1cm}
We consider  point-to-point communications where the transmitter has $\Ntx$ transmit ports (e.g., antennas), and the receiver has $\Nrx$ receive ports. We focus on multi-band transmission over $\Nbins$ subbands, each modeled as a linear \ac{awgn} channel. Letting $\myVec{y}[b] \in \mathbb{C}^{\Nrx}$ and $\myVec{s}[b] \in \mathbb{C}^{\Ntx}$ be the channel output and input at the $b$th subband, respectively, the channel input-output relationship is 
\begin{equation}
\label{eqn:systemModel}
  \myVec{y}[b] = \myMat{H}[b] \myVec{s}[b] + \myVec{w}[b].
\end{equation}
In \eqref{eqn:systemModel}, $\myMat{H}[b] \in \mathbb{C}^{\Nrx\times \Ntx}$ is the channel transfer function at the $b$th subband, while $\myVec{w}[b]\sim\mathcal{N}(\myVec{0},\sigma_w^2\myMat{I})$ is \ac{awgn}. 

The channel matrices $\{\myMat{H}[b]\}_{b=1}^{\Nbins}$ are dictated by varying parameters, e.g., the users' locations, which we model via the vector $\SoW$. In parameterized channels, the channel matrices are also affected by a set of $\Nparams$ parameters taking values in a feasible set $\mySet{P}$, denoted by $\Params \in \mySet{P}^{\Nparams}$, that are controllable. Accordingly, it is assumed that there exists a mapping $\mySet{M}_{\SoW}(\cdot)$ such that 
\begin{equation}
\label{eqn:Mapping}
    \big[\myMat{H}[1], \ldots, \myMat{H}[\Nbins] \big]= \mySet{M}_{\SoW}(\Params).
\end{equation}

The model \eqref{eqn:Mapping} is generic. As such, it accommodates common simplistic models for parametric channels in which $\mySet{M}$ is given in closed-form, such  as the cascaded model for \ac{ris}-aided communications~\cite{liu2019matrix}. Here, we focus on the realistic and challenging setup in which \eqref{eqn:Mapping} cannot be faithfully captured via a simple closed-form expression. Yet, one can still evaluate \eqref{eqn:Mapping} for a given setting of $\Params$ and $\SoW$, as in the following settings:
\begin{enumerate}[label={S\arabic*}]
    \item \label{itm:Simulator} The channel can be computed via a simulator based on, e.g., ray-tracing~\cite{vitucci2024efficient} or discrete dipole formalism~\cite{faqiri2022physfad}. 
    \item \label{itm:Measurement} The channel can be measured %for a given configuration 
    via, e.g., pilot signaling. 
\end{enumerate}

\subsection{Problem Formulation}
\label{ssec:problem formulation} 
The fact that $\Params$ are controllable allows us to set these values to achieve a desirable channel via \eqref{eqn:Mapping}. The channel figure-of-merit is %obtained using 
a differentiable mapping $\mySet{F}:\mathbb{C}^{\Nrx\times\Ntx\times\Nbins}\mapsto\mySet{R}$. An example is the achievable rate for uniform power allocation 
\begin{equation}
    \label{eqn:achieableRate}
    \mySet{F}\left(\{\myMat{H}[b]\}_{b=1}^B\right) = \frac{1}{\Nbins}\sum_{b=1}^{\Nbins}\log \left| \myMat{I} + \sigma_w^{-2}\cdot\myMat{H}[b]\myMat{H}^H[b] \right|. 
\end{equation}
Accordingly, our goal is to maximize $\mySet{F}(\cdot)$ through the controllable channel parameters. Namely, we seek to approach 

\begin{equation}
    \label{eqn:problem}
    \Params^{\star} = \argmax_{\Params \in \mySet{P}^{\Nparams}}\mathcal{F}(\Params;\SoW).  
\end{equation}

 %Our goal is to maximize $P$ through the change in the channel parameters: $$\hat{\varphi}=\argmax_\varphi\mathcal{F}\left(\mathcal{M}_\Psi\left(\varphi\right)\right)$$
 Tackling \eqref{eqn:problem} %under the generic model in \eqref{eqn:Mapping} 
 gives rise to several core challenges:
%There are several challenges we need to tackle:
\begin{enumerate}[label={C\arabic*}]
    \item \label{itm:compelx} The complex form of $\mySet{M}$ indicates that there is no closed-form solution for \eqref{eqn:problem}. Thus, one cannot derive $\Params^{\star}$ analytically, and must resort to numeric solutions.
    \item \label{itm:largeSearch} The search space $\mySet{P}^{\Nparams}$ can be very large, and possibly non-countable, with the vast majority of configurations leading to channels with only minor improvements in $\mySet{F}(\cdot)$.
    \item \label{itm:latency} The optimization procedure should be carried out rapidly, as  $\SoW$ can change frequently~\cite{raviv2023adaptive}, thus altering the mapping \eqref{eqn:Mapping}. 
%
    % \item \label{itm:Forwward} Capturing the forward measurement could be very complex and slow. This also prohibits the use of any method which needs to sample the environment many times.
    % \item \label{itm:convex} The problem is not necessarily convex, which means we might need to search in a different parameter space in order to achieve convexity. % I am trying to explain that through using NN we can achieve convexity..
\end{enumerate}
%
%Challenges \ref{itm:compelx}-\ref{itm:largeSearch} motivate using  descent methods~\cite[Ch. 9]{boyd2004convex} for optimizing parameterized channels. To do so efficiently, we seek to utilize first-order methods \cite[Ch. 9.3]{boyd2004convex}. % which can be computed using {\em differentiable digital twins}, as detailed in the following section. 

\subsection{Preliminaries: \ac{ald}}
Langevin dynamics accommodate a family of stochastic algorithms, typically used for sampling data from a distribution function $P(\myVec{x})$ using the (Stein) score function $\nabla_{\myVec{x}} \log P(\myVec{x})$~\cite{hyvarinen2005estimation}. This is achieved by iterating over~\cite{welling2011bayesian}
%
%We introduce the concept of (Stein) score~\cite{hyvarinen2005estimation}. Given a probability density function $p(x)$, the score function is defined as the gradient of the log probability density $\nabla_x\log p(x)$. The score function is a vector field that points toward the maximal growth for each point in the probability density function. \ac{sgm} train a score model and use it to generate samples from the distribution $p(x)$. This is usually done by applying the discrete Langevin Dynamics \ac{sde}(reference here-song and hermon).
 \begin{equation}
    \label{eqn:LangevinDynamics}
     \myVec{x}_t  =  \myVec{x}_{t-1}  +  \frac{\epsilon}{2}\nabla_{\myVec{x}}\log P(\myVec{x}_{t-1})  +  \sqrt{\epsilon}\myVec{z}_t,
 \end{equation}
 where $\epsilon>0$ is a small step size, $\myVec{z}_t$ is  sampled from a distribution $\mathcal{N}(\myVec{0},\myMat{I})$. 
 The \ac{ald} method is based on \eqref{eqn:LangevinDynamics}, while facilitating optimizing over complex distributions by, instead of  calculating the score of the clean distribution $P(\myVec{x})$, it uses the  distribution of $\tilde{\myVec{x}}=\myVec{x}+\myVec{v}$, denoted $P_{\sigma}(\tilde{\myVec{x}})$, where $\myVec{v}_t$ is \ac{awgn} with variance $\sigma^2$. The iterative procedure is given by
 \begin{equation}
    \label{eqn:AnnealedLangevinDynamics}
     \myVec{x}_t  =  \myVec{x}_{t-1}  +  \frac{\epsilon}{2}\nabla_{\tilde{\myVec{x}}}\log P_{\sigma_t}({\myVec{x}}_{t-1})  +  \sqrt{\epsilon}\myVec{z}_t,
 \end{equation}
 with the sequence $\{\sigma_t\}$ being hyperparameters. 
 This form of optimization is particularly useful for generating samples from distributions with sparse support~\cite{song2019generative}. %\Nir{please assert and cite. Should the argument be $\nabla_{\tilde{\myVec{x}}}\log P_{\sigma_t}(\tilde{\myVec{x}}_{t-1}) $ or $\nabla_{{\myVec{x}}}\log P_{\sigma_t}({\myVec{x}}_{t-1}) $?}

\section{Optimizing Programmable Channels}
\label{sec:Optimizing Programmable Channels}

%%% Rationale

\subsection{AI-Aided ALD for Channel Optimization}
%%% Overall algorithm
Our proposed approach views channel optimization under \ref{itm:compelx}-\ref{itm:latency} as {\em generating configurations} that yield  valid channels in the sense of $\mySet{F}(\cdot)$. This is achieved by formulating a surrogate probability field governed by a distribution of the form
\begin{equation}
    \label{eqn:ProbabilityFormulation}
    P(\Params|\SoW) \propto e^{\alpha \cdot \mathcal{F}\left(\mathcal{M}_{\SoW}\left(\Params\right)\right)}, \quad  \Params\in\mathcal{P}^{\Nparams},
\end{equation}
for some $\alpha > 0$.
Obviously, the desired $\Params^\star$ in \eqref{eqn:problem} is the most likely configuration in the sense of \eqref{eqn:ProbabilityFormulation}. By \ref{itm:largeSearch}, the probability field in \eqref{eqn:ProbabilityFormulation} is expected to be highly sparse, motivating tackling the sampling problem via \ac{ald} iterations as in \eqref{eqn:AnnealedLangevinDynamics}.  

To cope with the complex and possibly intractable form of $\mySet{M}$ (\ref{itm:compelx}), we leverage the fact that for the distribution of $\tilde{\Params}= \Params + \myVec{v}$ (where $\myVec{v}$ is \ac{awgn} as in \eqref{eqn:AnnealedLangevinDynamics}), it holds that \cite{raphan2011least}
% \Tomer{This reference used this method though this is not it's origin. The first paper to show this (although in a much altered form) is AN EMPIRICAL BAYES APPROACH TO STATISTICS from 1956 by herbert where he thanks tweedie for the idea and from my understanding this formula is often called tweedies formula. another closer form to what we show is presented in Least Squares Estimation Without Priors or Supervision}
\begin{equation}
    \nabla_{\tilde{\Params}}\log P_{\sigma}(\tilde{\Params}|\SoW) = \frac{1}{\sigma^2}\left(\mathbb{E}\left[\Params| \tilde{\Params}; \SoW,\sigma \right] -  \tilde{\Params}\right), 
\label{eqn:ScoreDenoiser}
\end{equation}
where the stochastic expectation is taken based on  \eqref{eqn:ProbabilityFormulation}. 

The formulation in \eqref{eqn:ScoreDenoiser} indicates that if one can estimate the conditional expectation $\mathbb{E}\left[\Params| \tilde{\Params}; \SoW,\sigma \right]$, then \ac{ald}-based channel optimization can be carried out without having to specify $\mySet{M}$. Accordingly, following the rationale used for generative \ac{ai} models in \cite{kadkhodaie2020solving}, we replace the conditional expectation used to evaluate the score in \eqref{eqn:ScoreDenoiser} with a \ac{dnn} parameterized by $\DNNParams$, whose mapping, denoted $D_{\DNNParams}(\cdot)$, maps a (noisy) parameters vector $\myVec{\phi}\in \mySet{P}^{\Nparams}$, along with $\SoW$ and noise level $\sigma$, into an estimate of 
\begin{equation}
\label{eqn:ScoreDenoiserApproximation}
   D_{\DNNParams}(\myVec{\phi}; \SoW, \sigma) \approx   \mathbb{E}\left[\Params| \tilde{\Params}=\myVec{\phi}; \SoW,\sigma \right].
\end{equation}
%\Nir{do we have a different \ac{dnn} for each iteration?}
 The resulting  optimization method is summarized as Algorithm~\ref{alg:ALD}.

  \begin{algorithm}
    \caption{ AI-aided \ac{ald} channel optimization}
    \label{alg:ALD} 
    \SetKwInOut{Initialization}{Init}
    \Initialization{\#time steps $T$ and iterations $K$; Trained \ac{dnn} $\DNNParams$;\\ 
      Hyperparameters $\epsilon$, $\{\sigma_t\}$;  Initial guess $\Params_0$ }
    \SetKwInOut{Input}{Input}
    \Input{Environment parameters $\SoW$}
    {
        \For{$t =  1, \ldots T$ }{%
            \For{$k =  1, \ldots K$ }{%
                Randomize $\myVec{z} \sim \mySet{N}(\myVec{0},\myMat{I})$\;
                 
                $\Params_k \! \leftarrow \! \Params_{k\!-\!1} \!+\! \frac{\epsilon}{2\sigma_t^2}\big(D_{\DNNParams}(\Params_{k\!-\! 1}; \SoW, \sigma_t)\!-\!\Params_{k\!-\!1} \big)$ 
                $\!+\! \sqrt{\epsilon}\myVec{z}$\;
            }
            $\Params_0\leftarrow\Params_K$
        }
    \KwRet{$\Params_K$}

  }
\end{algorithm}

\subsection{Training}
\label{ssec:training}
%\Nir{before the zero-order approximation, we should start with the loss function, because this is the interesting part. We should start by explaining why this is so different from how score based models are trained in generated AI, Now, for the training, the thing is that we want to in order to have \eqref{eqn:ScoreDenoiserApproximation} is: $(i)$ relate MMSE estimate to providing an output that is likely in the sense of the rate. We can do this by replacing the MMSE estimate with the MAP, for which we should get exactly what we are looking for; $(ii)$ if we stick with the MMSE interpretation, we should note that since the noise has variance $\sigma^2$ it should hold that $\mathbb{E}[\| D_{\DNNParams}(\myVec{\phi}; \SoW, \sigma) - \myVec{\phi} \|^2] = {\rm MMSE} + {\rm var}(\myVec{v})$. Below is how I think we should formulate it following $(i)$.}

\subsubsection{Loss Function}
The formulation of Algorithm~\ref{alg:ALD} is similar to the usage of \ac{ald} in generative \ac{ai}. However, we note two key differences between our channel optimization problem and, e.g., image generation: 
$(i)$ When training generative models, one typically has data of valid samples (e.g., natural images), while we cannot assume access to configurations known to generate "good" channels; 
$(ii)$ The key challenge in generative \ac{ai} is the lack of a prior distribution, while in our representation of channel optimization, we formulate a proxy distribution in \eqref{eqn:ProbabilityFormulation}.

A leading approach in diffusion models trains $D_{\myVec{\theta}}(\cdot)$ to denoise Gaussian perturbations~\cite{kadkhodaie2020solving}, 
% \Tomer{There is a leading approach in training score models this way and it is called Noise Conditional Score Network(NCSN), The use of denoisers is not used as much other than some }
leveraging the availability of valid samples and bypassing the need to impose a distribution, while approximating the \ac{mmse} estimate in \eqref{eqn:ScoreDenoiserApproximation}. Based on the above differences, we deviate from this existing training method, and train $D_{\myVec{\theta}}$ to approach the {\em \ac{map} estimate}, as a proxy for the \ac{mmse} estimate \eqref{eqn:ScoreDenoiserApproximation}.  Specifically, the \ac{map} rule naturally leverages \eqref{eqn:ProbabilityFormulation}, as stated in the following lemma:
\begin{lemma}
    \label{lem:MAP}
    The \ac{map} estimate of $\Params$ from the (noisy) realization $\tilde{\Params}=\myVec{\phi}$ for a fixed $\SoW$ and $\sigma$ is given by
    \begin{equation}
        \Params_{\rm MAP}(\myVec{\phi}) = \mathop{\arg \max}_{\Params \in \mySet{P}^{\Nparams}} \alpha \cdot \mathcal{F}\left(\mathcal{M}_{\SoW}\left(\Params\right)\right) - \frac{1}{2\myVec{\sigma}^2}\|\myVec{\phi} - \Params\|^2.
        \label{eqn:MAP}
    \end{equation}
\end{lemma}

\begin{IEEEproof}
    The lemma follows from  Bayes rule, as
    \begin{align}
        P(\Params|\tilde{\Params}; \SoW, \sigma) &\propto P(\tilde{\Params}|{\Params}; \SoW, \myVec{\sigma}) P({\Params}; \SoW, \myVec{\sigma}) \notag \\
        &\propto e^{\alpha \cdot \mathcal{F}\left(\mathcal{M}_{\SoW}\left(\Params\right)\right) - \frac{1}{2\myVec{\sigma}^2}\|\tilde{\Params} - \Params\|^2}.
        \label{eqn:posterior}
    \end{align}
    Thus, \eqref{eqn:MAP} is obtained by maximizing the (log) of \eqref{eqn:posterior}.
\end{IEEEproof} 
Based on Lemma~\ref{lem:MAP}, we formulate a loss function for training $\DNNParams$. Specifically, given a dataset of the form $\mySet{D} = \{\myVec{\phi}^{(i)}, \SoW^{(i)}, \myVec{\sigma}^{(i)}\}$, the empirical risk is given by 
\begin{align}
    \label{eqn:loss}
    \mySet{L}_{\mySet{D}}(\DNNParams) = \frac{1}{|\mySet{D}|}&\sum_{i=1}^{|\mySet{D}|} 
    \frac{\lambda}{(\myVec{\sigma}^{(i)})^2} \left\|\myVec{\phi}^{(i)} - D_{\DNNParams}(\myVec{\phi}^{(i)}; \SoW^{(i)}, \myVec{\sigma}^{(i)})\right\|^2 \notag \\
    &-    \mathcal{F}\left(\mathcal{M}_{\SoW}\left(D_{\DNNParams}(\myVec{\phi}^{(i)}; \SoW^{(i)}, \myVec{\sigma}^{(i)})\right)\right),
\end{align} 
where $\lambda > 0$ is a hyperparameter (corresponding to $1/2\alpha$ in \eqref{eqn:MAP}), 
The loss in \eqref{eqn:loss}, which is analytically obtained from Lemma~\ref{lem:MAP}, trains the \ac{dnn} to output the configuration that yields the best channel (in the sense of $\mySet{F}(\cdot)$), while encouraging this configuration to be in some proximity (dictated by $\sigma$) of the initial configuration. Accordingly, training using \eqref{eqn:loss} can be achieved without necessarily having a 'good' initial configuration, while leveraging the ability to assess a channel using $\mySet{F}(\cdot)$. 

%\Nir{what do you think? If you are fine with how I re-wrote things, please reformulate the remainder of the section accordingly. Note that our formulation is generic. There is no RIS or rate here, and instead of positions use state-of-the-world $\SoW$. }

%In the last section we have shown a way to generate optimal configuration given a trained denoiser, Though training a denoiser in the traditional way, by adding noise to clean data and training the denoiser in the removal of such noise, is impossible since we have no access to the "clean data" which for us would be optimal \ac{ris} configurations. This forces us to think of a new way to train.

\subsubsection{Computing Loss Gradients}
Training  according to \eqref{eqn:loss} using conventional first-order methods requires taking its gradient with respect to $\DNNParams$. The key challenge in doing so stems from the complex form of $\mySet{M}$ (\ref{itm:compelx}), which limits the ability to analytically take the gradient of the second term in \eqref{eqn:loss}. Particularly, to utilize backpropagation, one has to be able to compute the gradient $\nabla_{\Params}\mathcal{F}\left(\mathcal{M}_{\SoW}\left(\Params =D_{\DNNParams}(\myVec{\phi}; \SoW, \myVec{\sigma}) \right)\right)$. 

To enable gradient-based learning while coping with \ref{itm:compelx}, we propose to use {\em pseudo-gradients} during learning, computed via zero-order methods. We use stochastic unbiased  zero order gradient estimate via $2m$ measurements with $m<\Nparams$, for which
\begin{align}
    &\!\!\!\hat{\nabla}_{\Params}\mathcal{F}\left(\mathcal{M}_{\SoW}\left(\Params \right)\right)=\notag\\
    &\!\!\! \frac{\Nparams}{m} \sum_{j=1}^{m} \frac{\mathcal{F}\left(\mathcal{M}_\SoW\left(\Params \!+ \! \epsilon\cdot \myVec{u}_j \right)\right) \!- \! \mathcal{F}\left(\mathcal{M}_\SoW\left( \Params \! - \! \epsilon\cdot \myVec{u}_j \right)\right)}{2\epsilon}\myVec{u}_j,
    \label{eqn:zeroOrderGradientApprox}
\end{align}
where $\epsilon>0$ is a hyperparameter. By randomizing $\{\myVec{u}_j\}_{j=1}^m$  independently from a uniform distribution over the $\Nparams$-dimensional unit sphere, a stochastic gradient estimate is obtained~\cite{duchi2015optimal}. The pseudo gradients in \eqref{eqn:zeroOrderGradientApprox} are used to compute the gradient of the overall loss \eqref{eqn:loss} via backpropagation, enabling its adaptation with gradient-based learning. 

\subsubsection{Active Learning Based Training}
% OLD
% A crucial part of the success of the \ac{ald} algorithm is it's ability to learn the distribution of the data and noise with any power. This was achieved by training the model to predict the score function of the data with different levels of noise. Since we changed the training process to work only with random inputs which have no correlation to the output distribution we needed to find a another way of incorporating this distribution. To achieve this we saved the \ac{ris} configuration from the denoiser in conjunction with the transmitter location and noise level which the model used in a limited memory database. for each iteration we then choose whether to generate a new input or to take one at random from the distribution. The overall training algorithm is described in ~\ref{alg:trainingDenoiser}
% CHAT_GPT'D + personal editing
The \ac{map}-inspired loss in \eqref{eqn:loss} and the usage of pseudo gradients in \eqref{eqn:zeroOrderGradientApprox} enable the training of $\DNNParams$ from data. 
However, the annealing process in the \ac{ald} requires learning with a different input  distribution for each level of noise. We  tackle this by measuring channel realizations for different configurations $\Params$ and different settings $\SoW$ (e.g., different user locations) during training.
% However, as detailed above, we cannot assume prior access to "good" channels (as opposed to the training of generative models), but instead can measure channel realizations for different configurations $\Params$ and different settings $\SoW$ (e.g., different user locations) during training. 
Accordingly, we propose a training procedure based on active learning principles~\cite{be2019active}. On each iteration, the procedure has some probability $\gamma \in [0,1]$ of inspecting new initial configurations, and some probability $1-\gamma$ of refining previous configurations with lesser noise level $\sigma$. This  ensures that the model learns from a diverse yet structured distribution. 
The overall training algorithm is described in Algorithm~\ref{alg:trainingDenoiser}.

%A key factor in the success of the \ac{ald} algorithm is its ability to learn the distribution of data and noise at any power level. This is achieved by training the model to predict the score function at varying noise levels. Since we modified the training process to use only random, uncorrelated inputs, we needed a new way to incorporate the output distribution. To achieve this, we stored the output \ac{ris} configuration from the denoiser $\Params$, the state of the world $\SoW$, and the noise level $\sigma$ in a database. During training, each iteration either generates a new input or retrieves a previously stored sample to ensure the model learns from a diverse yet structured distribution.
%This process can also be applied in real time as a form of online learning to achieve crude ability to adapt to unkown environment changes although we admit to not testing this ability yet. \Tomer{As I am writing this sentence I think it should go to the discussion}
%The overall training algorithm is described in ~\ref{alg:trainingDenoiser}

%%%% overall training procedure
\begin{algorithm}
    \caption{Active Learning Based Training}
    \label{alg:trainingDenoiser}
    \SetKwInOut{Initialization}{Init}
    \Initialization{Initial weights $\DNNParams$; \\
        Learning rate $\eta$, number of iterations $I$; \\
        Hyperparameters  $\beta,\gamma\in [0,1]$.}%initial $\sigma$
    \SetKwInOut{Input}{Input} 
    \Input{Channel sampler $\mySet{M}_{\SoW}(\cdot)$; Settings $\{\SoW^{(i)}\}$ }
    Set $\mySet{D}\leftarrow\{(\NoisyParams^{(i)},\SoW^{(i)}, \myVec{\sigma}^{(i)})\}$ with random $\{\NoisyParams^{(i)}\}$\;
        \For{${\rm iter} = 1, \ldots, I$}{
            Randomize $v$ uniformly over $[0,1]$\;
            \uIf {$v>\gamma$}{
            %      Draw $\left\{ \NoisyParams,\SoW, \myVec{\sigma}\right\} \leftarrow \mathcal{D}_{mem}$\;
            % }
            % \Else{
                      $\mySet{D}\leftarrow\{(\NoisyParams^{(i)},\SoW^{(i)}, \myVec{\sigma}^{(i)})\}$ with random $\{\NoisyParams^{(i)}\}$\;
            }
            Compute  $\Params^{(i)} = D_\theta(\NoisyParams^{(i)};\SoW^{(i)}, \myVec{\sigma}^{(i)})$,  \;
            
            Compute pseudo gradients using \eqref{eqn:zeroOrderGradientApprox}\;
            
            Update  $\DNNParams\leftarrow \DNNParams - \eta \nabla_{\DNNParams}\mySet{L}_{\mySet{D}}(\DNNParams)$ according to \eqref{eqn:loss}\;
            
            Update $\myVec{\sigma}^{(i)} \leftarrow \beta \cdot \myVec{\sigma}^{(i)}$ and $\mySet{D}\leftarrow\{(\Params^{(i)},\SoW^{(i)}, \myVec{\sigma}^{(i)} )\}$\;
        }
    \SetKwInOut{Output}{Output}
    \Output{$\DNNParams$}
\end{algorithm}

%------------------------------------------------------------------------
%	Discussion
%------------------------------------------------------------------------
\subsection{Discussion}
Our proposed \ac{ai}-aided \ac{ald} channel optimizer is  designed to tune parameterized channels under \ref{itm:compelx}-\ref{itm:largeSearch}. It draws inspiration from the success of \ac{ald} in generative \ac{ai} in sampling from large sparse spaces (\ref{itm:largeSearch}), and alters its formulation to cast channel optimization as sampling. Our representation of the \ac{map}-inspired loss and the usage of pseudo gradients during learning exploit our ability to analytically asses a given channel and enable gradient based learning that copes with \ref{itm:compelx}. While one can potentially use pseudo gradients for channel optimization, doing so during inference entails notable latency due to the need to repeatedly measure the channel. Our training method requires multiple measurements (for computing the pseudo gradients) in offline training, while upon a given $\SoW$, the selection of $\Params$ is rapid and does not involve any additional channel measurements (meeting \ref{itm:latency}).

The proposed methodology alters the deign of diffusion-type \acp{dnn} by a \ac{map}-oriented objective and the usage of pseudo gradients. It notably differs from existing application of diffusion-type \acp{dnn} for optimization purposes, which to date has focused mostly on inverse problems while preserving the conventional generative formulation. As such, our approach  is expected to be useful in various other optimization domains beyond that of parameterized channels.  We leave such extensions for future work.

\begin{figure}
    \centering
    \vspace{-0.15cm}
    \includegraphics[width=0.78\linewidth]{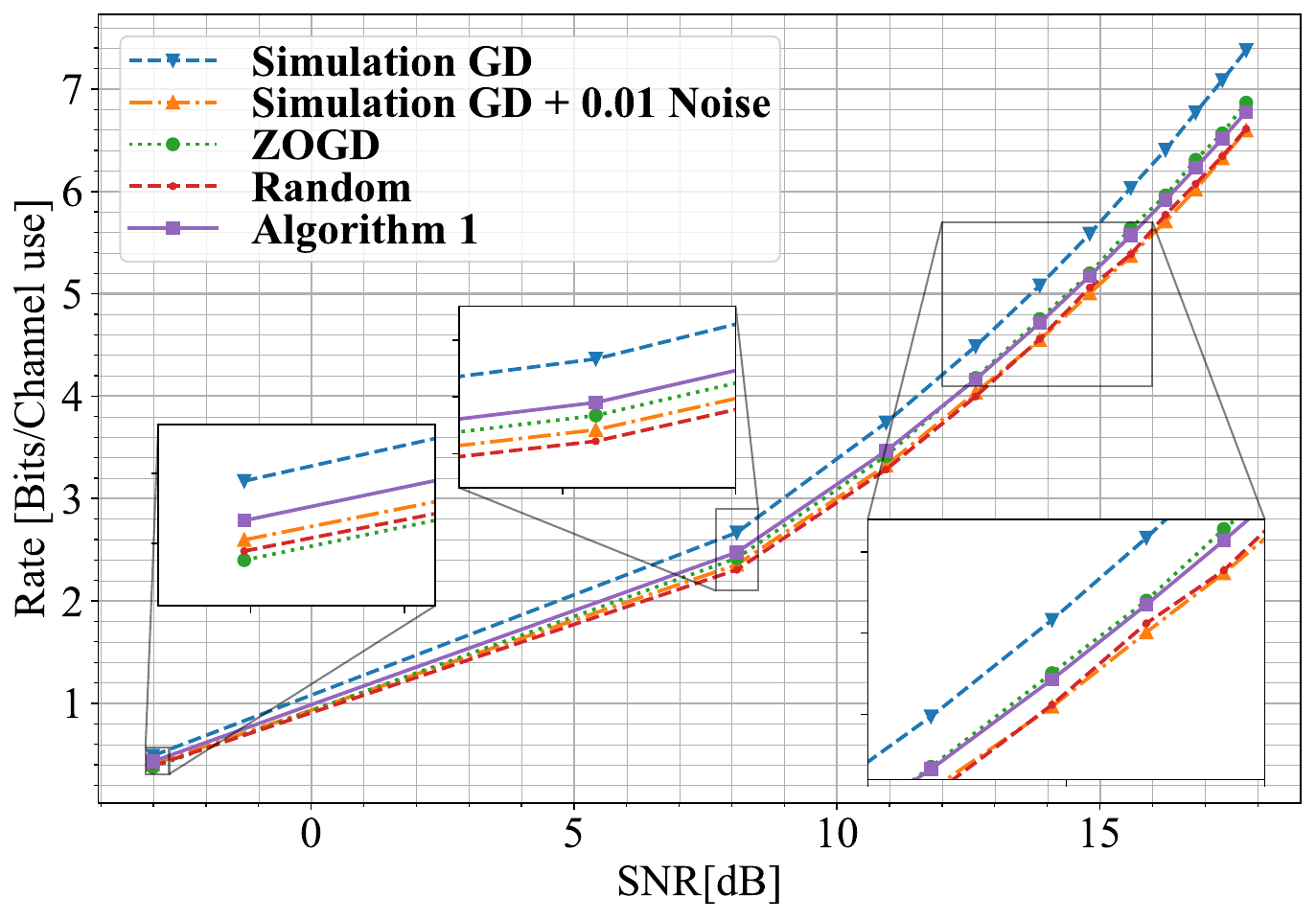}
    \caption{Achievable rate versus \acs{snr}.}
    \label{fig:SNR Place Holder}
\end{figure}

%------------------------------------------------------------------------
%	Numerical Results
%------------------------------------------------------------------------
%\vspace{-0.1cm}
\section{Numerical Results}
\label{sec:simulations}
\begin{figure}
    \centering
    \includegraphics[width=1\linewidth]{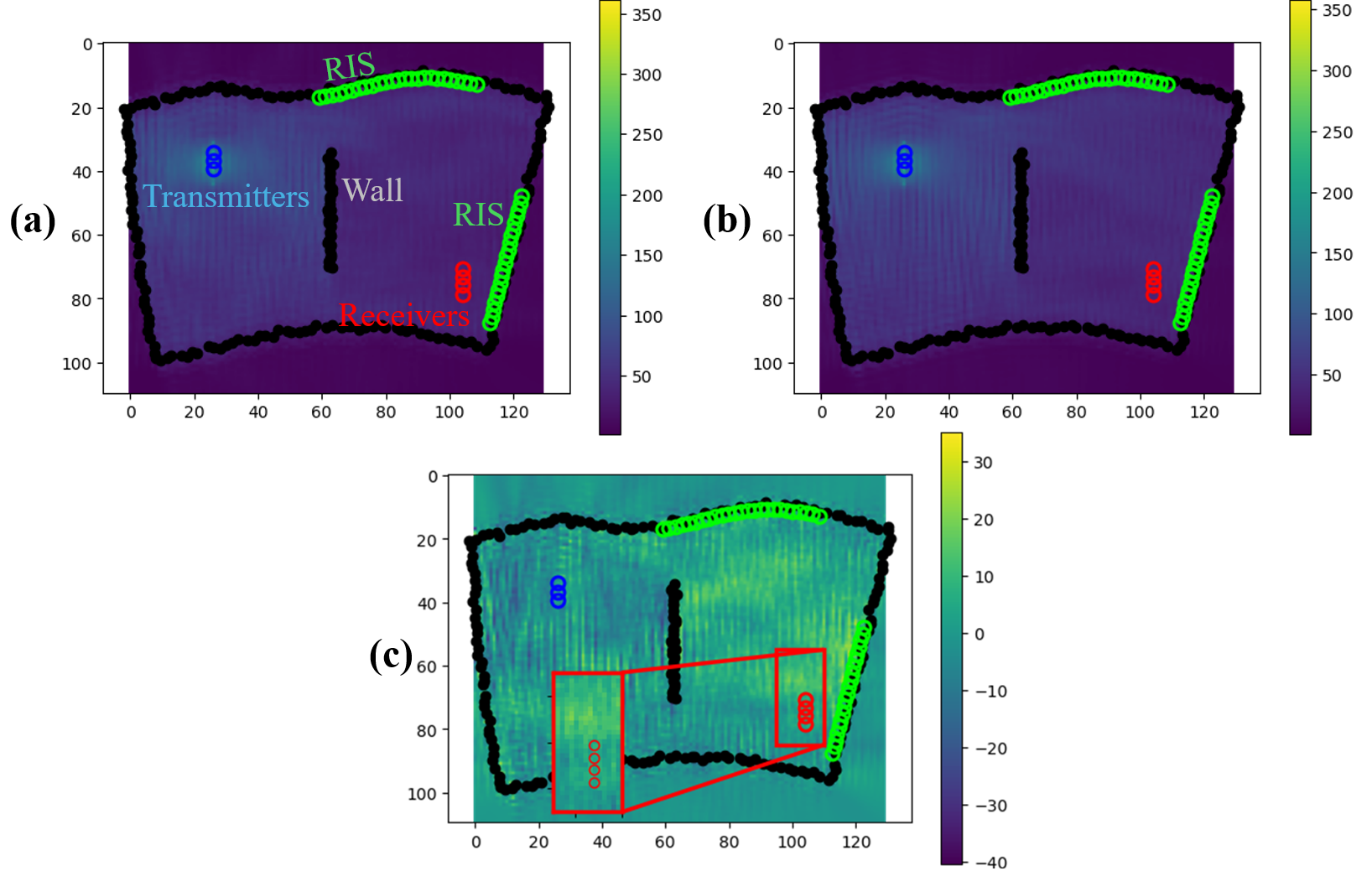}
    \caption{Achievable rate heatmap  $(a)$ without and $(b)$ with  optimizing via Algorithm~\ref{alg:ALD}, and $(c)$  difference between the rate heatmaps} %\textcolor{red}{I suspect that the horizontal and vertical scales are not the same. Also, can we increase the font size of the axis labels?}}
    \label{fig:spatial capacity}
\end{figure}

%\vspace{-0.1cm}
% General settings -  ris setting, goal, transmitors, receivers, number of panels and number of elements.
Our numerical study optimizes \acp{ris} in rich scattering settings\footnote{The source code and all hyperparameters used are available online at \url{https://github.com/tomer1203/Generative_RIS_Optimization}}. 
We use PhysFad~\cite{faqiri2022physfad} to generate channels  with $\Ntx=3$ transmit antennas and $\Nrx=4$ receive antennas. There is no line-of-sight (see  Fig.~\ref{fig:spatial capacity}), and there are two frequency selective \acp{ris} with  45 elements and $\Nparams=135$ configurable parameters (see \cite{katsanos2022wideband}). 
The parameters $\SoW$ dictate the localization of the transmitter antennas, which is randomized in each trial. 
%We use the transmitters locations as an example for ~\ref{itm:latency} of a dynamic \textit{'state of the world'} $\SoW$ which could change frequently.
The objective is the achievable rate in \eqref{eqn:achieableRate}.

To implement the channel optimizer of Algorithm~\ref{alg:ALD}, we employ a \ac{dnn} for $D_{\DNNParams}$ composed of six fully-connected layers,  ReLU activations, and sigmoid output layer. We train the \ac{dnn}  using Algorithm~\ref{alg:trainingDenoiser} for 180 iterations (with \ac{snr} of $1/\sigma_w^2=1$). The training hyperparameters are $\beta=0.9$, $\gamma=0.25$, and  $\eta = 0.0005$.
We compare Algorithm~\ref{alg:ALD} to two alternative methods for optimizing channel parameters: $(i)$ {\em ZOGD}: a measurement-based optimizer that tunes $\Params$ using zero-order gradient descent for channel optimization (as opposed to our method that uses zero-order gradients during training), with $2m=8$ points;
$(ii)$ {\bf Simulator}: an optimizer that has apriori access to a simulator (using the simulator of \cite{faqiri2022physfad}), and uses it for gradient-based optimization. When the simulator faithfully describes the environment, this approach represents an upper bound on the achievable performance. We thus consider both a setting where the channel is indeed generated with the simulator, as well as when the simulator is a approximation of the resulting channel, which we model by having the channel be a noisy version of the one produced by the simulator, obtained  by adding  a very small noise (of variance not surpassing $10^{-2}$ of the noise free channel).

\begin{figure}
    \centering
    \vspace{-0.15cm}
    \includegraphics[width=0.78\linewidth]{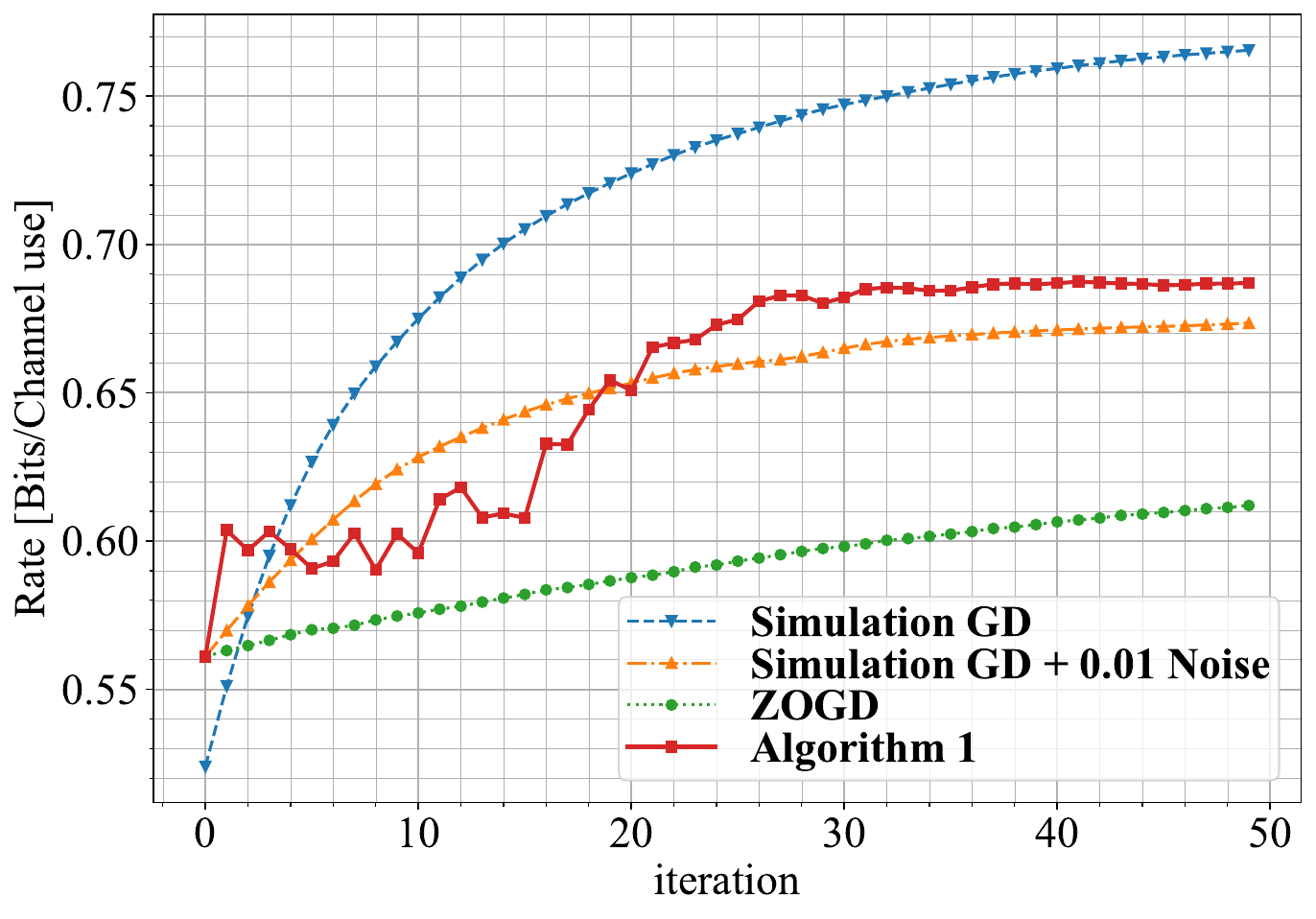}
    \caption{Achievable rate per iteration.}
    \label{fig:Rate Per Iteration}
\end{figure}

In Fig.~\ref{fig:SNR Place Holder} we present the rate of the considered methods when they run until convergence, versus \ac{snr}. For comparison, we also evaluate non-gradient based random search. 
Observing Fig.~\ref{fig:SNR Place Holder}, we note that
while the differentiable simulator with complete knowledge of the environment achieves the best rate. Though when even a small level of noise is added to the environment then the rate degrades significantly. We can also observe that using Algorithm~\ref{alg:ALD} produces comparatively good results even surpassing  ZOGD which aided it's training for some \ac{snr} values. 
The differences between the algorithms in  convergence rate are highlighted in Fig.~\ref{fig:Rate Per Iteration}. There, each method runs for 50 iterations (Algorithm~\ref{alg:ALD} uses $T=10$ time steps and $K=5$ iterations per time step). Algorithm~\ref{alg:ALD} returns the best performance after 50 iterations (except for the full knowledge simulator). The ZOGD improves too slowly to allow any improvement in this setting.

% when one has access to a differentiable simulator that faithfully captures the channel, using it for first-order optimization yields the best configuration. 
% Its gradient computation is more lengthy than using \ac{dnn} digital twins, and when restricted in latency, both approaches yield similar rates. The \ac{dnn} twin is most beneficial in low \acp{snr} (where it was trained), while generalizing to high \acp{snr}. Zero-order optimization  works well when allowed to run until convergence, yet under latency limits, it does not surpass random search.
%run every algorithm with different \ac{snr} values to stabilization in rate. We can see that for all \ac{snr}'s we get the highest achieved rate through the simulator optimization. Since we are also interested in fast convergence rates we give a fair comparison to the \ac{dnn} algorithm by limiting all other algorithms optimization time to be less than that of the \ac{dnn}-aided digital twin. In the limited run time setting we can see that the \ac{dnn}-aided model achieves similar results to the simulator digital twin and better results than both other methods. Another interesting observation is that the \ac{dnn} model gives comparatively better results for lower SNR values, this phenomena might be attributed to the fact that the training was performed only on these \ac{snr} ranges.

\begin{table}
\centering
\vspace{0.3cm}
\caption{Latency evaluation}
%\begin{adjustbox}{width=\columnwidth} 
\begin{tabular}{|p{3.4cm}|p{1.3cm}|p{2.6cm}|}
\hline
Method                        & Latency [s] &  Rate [bits/channel use]  \\ \hline\hline   
Simulator perfect knowledge   &  125.19 & \textbf{0.453}          \\ \hline     
Simulator imperfect knowledge &  119.56 & 0.372                   \\ \hline     
Zero-order gradient descent   &  473.57 & 0.369                   \\ \hline     
Algorithm~\ref{alg:ALD}          &  \textbf{0.145}  & \textbf{0.414}                   \\ \hline     
\end{tabular}
\label{tbl:NumComp}
%\end{adjustbox}
\vspace{-0.3cm}
\end{table}

% Numeric table evaluation discussion
To capitalize on the latency-performance tradeoff, we report in Table~\ref{tbl:NumComp} the rate achieved per latency for the same test used for Fig.~\ref{fig:Rate Per Iteration}. All methods are computed on the same computer (i9-13900H,32G RAM, RTX 4050). While these values can be  reduced by using hardware accelerators, they indicate on the relative expected latency. The results presented are the latency required to optimize $8$ configurations  and the mean rate achieved on the batch. We observe in Table~\ref{tbl:NumComp} that Algorithm~\ref{alg:ALD}  yields configurations with rates within a minor gap of that achieved using a simulator with perfect knowledge. However, in terms of latency, Algorithm~\ref{alg:ALD} shows more than three orders of magnitude of improvement compared to all other methods.

% \begin{figure}
%     \centering
%     \includegraphics[width=1\linewidth]{SNR.png}
%     \caption{\color{blue}Graph curve will be smoother after I finish running with a longer sample}
%     \label{fig:SNR_graph}
% \end{figure}
% \begin{figure}
%     \centering
%     \includegraphics[width=1\linewidth]{SNR_DB.png}
%     \caption{\color{blue}Actually in DB. after looking online this seems to match with the theory. What do you think? which one would be more clear to a communication researcher?(ignore the x-x-ticks in this graph) I think it is more readable without the DB but it might be very unorthodox.} 
%     \label{fig:SNR in DB}
% \end{figure}

% \begin{figure}
%     \centering
%     \includegraphics[width=1\linewidth]{room_configurations_comparison.png}
%     \caption{spatial capacity comparison between the room before and after the optimization process.}
%     \label{fig:spatial capacity MSE}
% \end{figure}

We conclude by visually assessing how Algorithm~\ref{alg:ALD} optimizes parameterized channels.  Fig.~\ref{fig:spatial capacity}  visualizes how such  optimization generates favorable conditions in the  location of the receive antennas. The gap between an unoptimized setting and the optimized one is highlighted in Fig.~\ref{fig:spatial capacity}(c), noting a clear rate growth in the  receivers area compared to the rest of the room.

%------------------------------------------------------------------------
%	Conclusions
%------------------------------------------------------------------------
%\vspace{-0.1cm}
\section{Conclusions}
\label{sec:conclusions}
%\vspace{-0.1cm}
We proposed an \ac{ai}-aided algorithm for optimizing parameterized channels. Our method is based on altering \ac{ald}, commonly used for diffusion generative models, to be based on \ac{map}-type models. We proposed a dedicated learning method that combines active learning with pseudo gradients. We numerically show that the proposed approach enables the tuning of useful \ac{ris} configurations in challenging settings with notably reduced latency.

%\vspace{-0.1cm}
\bibliographystyle{IEEEtran}
\bibliography{IEEEabrv,refs}

\begin{thebibliography}{10}
\providecommand{\url}[1]{#1}
\csname url@samestyle\endcsname
\providecommand{\newblock}{\relax}
\providecommand{\bibinfo}[2]{#2}
\providecommand{\BIBentrySTDinterwordspacing}{\spaceskip=0pt\relax}
\providecommand{\BIBentryALTinterwordstretchfactor}{4}
\providecommand{\BIBentryALTinterwordspacing}{\spaceskip=\fontdimen2\font plus
\BIBentryALTinterwordstretchfactor\fontdimen3\font minus \fontdimen4\font\relax}
\providecommand{\BIBforeignlanguage}[2]{{%
\expandafter\ifx\csname l@#1\endcsname\relax
\typeout{** WARNING: IEEEtran.bst: No hyphenation pattern has been}%
\typeout{** loaded for the language `#1'. Using the pattern for}%
\typeout{** the default language instead.}%
\else
\language=\csname l@#1\endcsname
\fi
#2}}
\providecommand{\BIBdecl}{\relax}
\BIBdecl

\bibitem{giordani2020toward}
M.~Giordani \emph{et~al.}, ``{Toward 6G networks: Use cases and technologies},'' \emph{{IEEE} Commun. Mag.}, vol.~58, no.~3, pp. 55--61, 2020.

\bibitem{alexandropoulos2021reconfigurable}
G.~C. Alexandropoulos \emph{et~al.}, ``Reconfigurable intelligent surfaces for rich scattering wireless communications: Recent experiments, challenges, and opportunities,'' \emph{{IEEE} Commun. Mag.}, vol.~59, no.~6, pp. 28--34, 2021.

\bibitem{molisch2017hybrid}
A.~F. Molisch \emph{et~al.}, ``Hybrid beamforming for massive {MIMO}: A survey,'' \emph{{IEEE} Commun. Mag.}, vol.~55, no.~9, pp. 134--141, 2017.

\bibitem{shlezinger2021dynamic}
N.~Shlezinger \emph{et~al.}, ``Dynamic metasurface antennas for {6G} extreme massive {MIMO} communications,'' \emph{{IEEE} Wireless Commun.}, vol.~28, no.~2, pp. 106--113, 2021.

\bibitem{gabay2023leaky}
Y.~Gabay \emph{et~al.}, ``Leaky waveguide antennas for downlink wideband {THz} communications,'' in \emph{Proc. IEEE ICASSP}, 2024.

\bibitem{faisal2022machine}
K.~Faisal and W.~Choi, ``Machine learning approaches for reconfigurable intelligent surfaces: A survey,'' \emph{{IEEE} Access}, vol.~10, pp. 27\,343--27\,367, 2022.

\bibitem{shlezinger2023ai}
N.~Shlezinger \emph{et~al.}, ``Artificial intelligence-empowered hybrid multiple-input/multiple-output beamforming: Learning to optimize for high-throughput scalable {MIMO},'' \emph{{IEEE} Veh. Technol. Mag.}, vol.~19, no.~3, pp. 58--67, 2024.

\bibitem{strinati2021reconfigurable}
E.~Calvanese~Strinati \emph{et~al.}, ``Reconfigurable, intelligent, and sustainable wireless environments for 6{G} smart connectivity,'' \emph{{IEEE} Commun. Mag.}, vol.~59, no.~10, pp. 99--105, 2021.

\bibitem{boyd2004convex}
S.~P. Boyd and L.~Vandenberghe, \emph{Convex optimization}.\hskip 1em plus 0.5em minus 0.4em\relax Cambridge university press, 2004.

\bibitem{lavi2023learn}
O.~Lavi and N.~Shlezinger, ``Learn to rapidly and robustly optimize hybrid precoding,'' \emph{{IEEE} Trans. Commun.}, vol.~71, no.~10, pp. 5814--5830, 2023.

\bibitem{liu2019matrix}
H.~Liu \emph{et~al.}, ``Matrix-calibration-based cascaded channel estimation for reconfigurable intelligent surface assisted multiuser {MIMO},'' \emph{{IEEE} J. Sel. Areas Commun.}, vol.~38, no.~11, pp. 2621--2636, 2020.

\bibitem{rabault2024tacit}
A.~Rabault \emph{et~al.}, ``On the tacit linearity assumption in common cascaded models of {RIS}-parametrized wireless channels,'' \emph{{IEEE} Trans. Wireless Commun.}, vol.~23, no.~8, pp. 10\,001--10\,014, 2024.

\bibitem{swindlehurst2021channel}
A.~L. Swindlehurst \emph{et~al.}, ``Channel estimation with reconfigurable intelligent surfaces-- {A} general framework,'' \emph{Proc. {IEEE}}, vol. 110, no.~9, pp. 1312--1338, 2022.

\bibitem{alexandropoulos2023hybrid}
G.~C. Alexandropoulos \emph{et~al.}, ``Hybrid reconfigurable intelligent metasurfaces: Enabling simultaneous tunable reflections and sensing for 6{G} wireless communications,'' \emph{{IEEE} Veh. Technol. Mag.}, vol.~19, no.~1, pp. 75--84, 2024.

\bibitem{vitucci2024efficient}
E.~M. Vitucci \emph{et~al.}, ``An efficient ray-based modeling approach for scattering from reconfigurable intelligent surfaces,'' \emph{{IEEE} Trans. Antennas Propag.}, vol.~72, no.~3, pp. 2673--2685, 2024.

\bibitem{faqiri2022physfad}
R.~Faqiri \emph{et~al.}, ``Phys{F}ad: Physics-based end-to-end channel modeling of {RIS}-parametrized environments with adjustable fading,'' \emph{{IEEE} Trans. Wireless Commun.}, vol.~22, no.~1, pp. 580--595, 2022.

\bibitem{wang2021jointly}
L.~Wang \emph{et~al.}, ``Jointly learned symbol detection and signal reflection in {RIS}-aided multi-user {MIMO} systems,'' in \emph{Asilomar Signals Syst.}, 2021.

\bibitem{khan2022digital}
L.~U. Khan \emph{et~al.}, ``Digital-twin-enabled 6{G}: Vision, architectural trends, and future directions,'' \emph{{IEEE} Commun. Mag.}, vol.~60, pp. 74--80, 2022.

\bibitem{van2024generative}
N.~Van~Huynh \emph{et~al.}, ``Generative {AI} for physical layer communications: A survey,'' \emph{{IEEE} Trans. on Cogn. Commun. Netw.}, vol.~10, no.~3, pp. 706--728, 2024.

\bibitem{alkhateeb2023real}
A.~Alkhateeb \emph{et~al.}, ``Real-time digital twins: Vision and research directions for 6{G} and beyond,'' \emph{{IEEE} Commun. Mag.}, vol.~61, no.~11, pp. 128--134, 2023.

\bibitem{li2023learnable}
B.~Li \emph{et~al.}, ``Learnable digital twin for efficient wireless network evaluation,'' in \emph{Proc. IEEE MILCOM}, 2023.

\bibitem{lee2024generating}
T.~Lee, J.~Park, H.~Kim, and J.~G. Andrews, ``Generating high dimensional user-specific wireless channels using diffusion models,'' \emph{arXiv preprint arXiv:2409.03924}, 2024.

\bibitem{zhang2025decision}
J.~Zhang \emph{et~al.}, ``Decision transformers for {RIS}-assisted systems with diffusion model-based channel acquisition,'' \emph{arXiv preprint arXiv:2501.08007}, 2025.

\bibitem{tong2024diffusion}
W.~Tong \emph{et~al.}, ``Diffusion model-based channel estimation for {RIS}-aided communication systems,'' \emph{{IEEE} Wireless Commun. Lett.}, vol.~13, no.~9, pp. 2586--2590, 2024.

\bibitem{song2019generative}
Y.~Song and S.~Ermon, ``Generative modeling by estimating gradients of the data distribution,'' \emph{Advances in neural information processing systems}, vol.~32, 2019.

\bibitem{kawar2022denoising}
B.~Kawar \emph{et~al.}, ``Denoising diffusion restoration models,'' \emph{Advances in Neural Information Processing Systems}, vol.~35, pp. 23\,593--23\,606, 2022.

\bibitem{shlezinger2022model}
N.~Shlezinger \emph{et~al.}, ``Model-based deep learning: On the intersection of deep learning and optimization,'' \emph{{IEEE} Access}, vol.~10, pp. 115\,384--115\,398, 2022.

\bibitem{duchi2015optimal}
J.~C. Duchi \emph{et~al.}, ``Optimal rates for zero-order convex optimization: The power of two function evaluations,'' \emph{{IEEE} Trans. Inf. Theory}, vol.~61, no.~5, pp. 2788--2806, 2015.

\bibitem{stylianopoulos2022deep}
K.~Stylianopoulos \emph{et~al.}, ``Deep-learning-assisted configuration of reconfigurable intelligent surfaces in dynamic rich-scattering environments,'' in \emph{Proc. IEEE ICASSP}, 2022.

\bibitem{be2019active}
I.~Be’ery, N.~Raviv, T.~Raviv, and Y.~Be’ery, ``Active deep decoding of linear codes,'' \emph{{IEEE} Trans. Commun.}, vol.~68, no.~2, pp. 728--736, 2019.

\bibitem{raviv2023adaptive}
T.~Raviv \emph{et~al.}, ``Adaptive and flexible model-based {AI} for deep receivers in dynamic channels,'' \emph{{IEEE} Wireless Commun.}, vol.~31, no.~4, pp. 163--169, 2024.

\bibitem{hyvarinen2005estimation}
A.~Hyv{\"a}rinen and P.~Dayan, ``Estimation of non-normalized statistical models by score matching.'' \emph{Journal of Machine Learning Research}, vol.~6, no.~4, 2005.

\bibitem{welling2011bayesian}
M.~Welling and Y.~W. Teh, ``Bayesian learning via stochastic gradient {L}angevin dynamics,'' in \emph{International Conference on Machine Learning}.\hskip 1em plus 0.5em minus 0.4em\relax Citeseer, 2011, pp. 681--688.

\bibitem{raphan2011least}
M.~Raphan and E.~P. Simoncelli, ``Least squares estimation without priors or supervision,'' \emph{Neural computation}, vol.~23, no.~2, pp. 374--420, 2011.

\bibitem{kadkhodaie2020solving}
Z.~Kadkhodaie and E.~P. Simoncelli, ``Solving linear inverse problems using the prior implicit in a denoiser,'' \emph{arXiv preprint arXiv:2007.13640}, 2020.

\bibitem{katsanos2022wideband}
K.~D. Katsanos \emph{et~al.}, ``Wideband multi-user {MIMO} communications with frequency selective {RIS}s: Element response modeling and sum-rate maximization,'' in \emph{Proc. IEEE ICC}, 2022.

\end{thebibliography}
\end{document}